\documentclass[prl,showpacs,twocolumn]{revtex4}

\usepackage{amsmath}
\usepackage{amssymb}

\usepackage{color}
\usepackage{graphicx}
\usepackage[english]{babel}

\begin{document}

\title{Density of states in spin-valve structure with superconducting electrodes}

\author{D. Gusakova, A. Vedyayev, O. Kotel'nikova}
\address{
Magnetism Department, Faculty of Physics, M.V.Lomonosov Moscow
State University, Moscow 119992,Russia}

\begin{abstract}
Energy variation of the density of states (DOS) has been
calculated in the
superconductor/ferromagnet/ferromagnet/superconductor structure
(SFFS) in the frame of Gorkov equations taking into account the
s-d electron scattering in the ferromagnetic layers. DOS behavior
is presented for the antiparallel and parallel magnetic moments
alignment of two adjacent F layers. The cases of small and large
values of exchange ferromagnetic field are discussed.
\end{abstract}


\maketitle

\subsection{I. INTRODUCTION}

 The possibility of the potential application of multilayers
with the superconducting junctions stimulate an interest to the
study of such a structures. The effect of switching between "0"
and "$\pi$" states in
superconductor/ferromagnet/ferromagnet/superconductor (SFFS)
structures
 makes them useful for the microelectronic engineering.
Practically it is possible to change the mutual orientation of the
thin ferromagnetic layers magnetization by applying the external
magnetic field. Two ferromagnetic layers separated by a
nonmagnetic spacer FNF are known as "spin-valve". The combined
SFFS structure with superconducting electrodes gives more
interesting results. In the parallel configuration for the some
ferromagnetic thickness values the Josephson current takes a zero
value. At the same time for the antiparallel alignment of
magnetizations the Josephson current has a non zero value. Thus,
by selecting a necessary ferromagnetic layer thickness we can
govern the superconducting current flow.

The presence of two ferromagnetic layers with the dirigible
magnetization alignment let study two opposite cases which have
their peculiarities. For the parallel magnetization alignment it
is the reversion of the Josephson current sign ($\pi$ state
junction) that has been theoretically described \cite{Buzdin1982}
and experimentally observed \cite{Kontos2001,Ryazanov2001}, and
for the antiparallel case the possible current divergence.

Several authors involved in the studying of SFIFS structures (I is
the insulating barier) touched upon a question of the possible
enhancement of the critical Josephson current for the antiparallel
magnetization alignment of the neighboring F layers under some
conditions. In \cite{BergeretLett2001} it was found that in the
case of the antiparallel orientation the critical current
increases at low temperatures with increasing exchange field $h$
and at zero temperature has a singularity when $h$ equals to the
superconducting energy gap $\Delta$. Authors of \cite{Koshina2001}
also predict the critical current enhancement for the antiparallel
alignment in some interval of exchange field values. The
possibility of the critical current enhancement by the exchange
field in SFIFS junctions with thin F layers with antiparallel
magnetization directions was discussed in the regimes os small S
layer thickness \cite{BergeretLett2001} and bulk S electrodes
\cite{Koshina2001}. Golubov et al. \cite{Golubov2002} assumed that
 the case of thin S layers \cite{BergeretLett2001}
was studied in the frame of an idealized model in the tunnelling
limit, which leads to a divergency of the critical current at the
zero temperature, and for the bulk S case \cite{Koshina2001} an
approximate method was used, so that a part of the results was
obtained beyond its applicability range. Golubov et al.
\cite{Golubov2002} explain the current enhancement by the
singularity in density of states (DOS) which is shifted to the
Fermi energy level. It happens when the effective energy shift in
the ferromagnets due to the exchange field becomes equal to a
local value of the effective energy gap induced into F layers.
They note that in the models with DOS of the BCS type this leads
to a logarithmic divergency of the critical current in the
antiparallel case at zero temperature.

First of all, it should be emphasized that in most theoretical
papers the SFFS structures were studied in the so-called dirty
limit and for low energies, that is using the quasiclassical
Usadel equations in the context of one-band model of ferromagnetic
metal. Such approach does not take into account the s-d scattering
of electrons which is the main scattering mechanism responsible
for the kinetic properties of 3d-metals and their conductivity.
Evidently, the limitation connected with the Usadel equations may
be avoided by solving the full Gorkov equations which in addition
take into account the s-d scattering of conducting electrons.

In the paper \cite{Pugach2004} the Josephson current in SFFS
structure had been calculated in the frame of Gorkov equations
taking into account the two-band model of a ferromagnet with
conducting s electrons and almost localized d electrons. S layers
were assumed as a simple s-wave superconductors. For the
antiparallel configuration the Josephson current has no
singularity at zero temperature. Consequently, from the physical
point of view the results derived in
\cite{BergeretLett2001,Koshina2001,Golubov2002} stay unclear and
this question requires a more careful study. For that reason in
this paper we calculate and discuss DOS for both (parallel and
antiparallel) cases in the frame of Gorkov equations taking into
account the s-d scattering of electrons. As it will be shown  DOS
does not demonstrate any divergence that makes us conclude that
s-d scattering effectively destroys the BCS correlation, hence,
the Josephson current enhancement is suppressed.

\subsection{II. GORKOV EQUATIONS}

  We consider the SFFS structure which
consists of two semi-infinite superconducting electrodes S
separated by two thin ferromagnetic layers F with the interfaces
perpendicular to the $z$ axis. Although the Green function method
can be applied in more general case, for simplicity we assume the
F and S materials to be equivalent on both sides of structure, and
both F layers have the same thickness $a$ which is supposed to be
much smaller than the in-plane dimension of the structure. The S/F
interfaces are placed at $z=\pm a$. In the mixed
$(\overrightarrow{k},z)$ representation ($\overrightarrow{k}$  is
the quasi-momentum component in the $XY$ plane of layers) the
Gorkov equations for the normal $G_{\uparrow\uparrow}^{ss}$ and
anomalous $F_{\downarrow\uparrow}^{ss}$ Green functions have the
following form:

\begin{widetext}
\begin{equation}\label{02}
   \begin{array}{l}
     \left[ i\omega + \frac{1}{2m} \left( \frac{\partial^2}{\partial z^2}-k^2 \right)+
     \varepsilon_F(z)+h(z)-x_0(z)\gamma^2 G_{\uparrow\uparrow}^{dd}(z,z,\omega) \right]
     G_{\uparrow\uparrow}^{ss}(z,z',\omega)+\Delta(z)F_{\downarrow\uparrow}^{ss}(z,z',\omega)=
     \delta(z-z'), \\
      \left[ i\omega - \frac{1}{2m} \left( \frac{\partial^2}{\partial z^2}-k^2
      \right)-
     \varepsilon_F(z)+h(z)-x_0(z)\gamma^2 G_{\downarrow\downarrow}^{dd}(z,z,\omega) \right]
     F_{\downarrow\uparrow}^{ss}(z,z',\omega)+\Delta^{\ast}(z)G_{\uparrow\uparrow}^{ss}(z,z',\omega)=
     0, \\
   \end{array}
\end{equation}
\end{widetext}
where $\omega=\pi T(2n+1)$ is the Matsubara frequency at
temperature $T$,
 $\varepsilon_F(z)=\varepsilon^s[\theta(-a-z)+\theta(z-a)]+\varepsilon^f
 \theta(z+a)\theta(a-z)$, $\varepsilon^{s(f)}$
  the Fermi
 energy of the superconductor (ferromagnet), $\theta(z)$ the step
 function ($\theta(z)=1$, if $z\geq0$, and $\theta(z)=0$, if $z<0$), $\hbar=k_B=1$.
We consider that the Cooper pairing is absent in the F layer and
the superconducting order parameter $\Delta$ is equal to zero in
the ferromagnet, $\Delta(z)=\Delta
exp(i\varphi_1)\theta(-a-z)+\Delta exp(i\varphi_2)\theta(z+a)$
($\varphi_{1,2}$ is the order parameter phase in the left and
right superconducting electrodes, correspondingly). The exchange
magnetic field in the F layer:
  $h(z)=h\theta(z+a)\theta(-z)\pm h \theta(z)\theta(a-z)$, $h>0$,
  sign $(+)$ corresponds to the parallel alignment of F layer
  magnetizations and $(-)$ sign
  should be taken for the antiparallel configuration.
 The s-d scattering is
described in the framework of the fist Born approximation with
impurity concentration $x_0(z)=x_0 \theta(z+a)\theta(a-z)$ and
impurity potential value $\gamma$.
$G_{\uparrow\uparrow(\downarrow\downarrow)}^{dd}(z,z,\omega)$ are
the diagonal Green functions of localized d electrons. Here we
neglect the s-s electron scattering as it is small in comparison
with the s-d scattering which are the main mechanism destroying
induced superconductivity in the F layers.

In this case the Gorkov equations become linear and the solutions
represent a set of plane waves with wave vectors:

\begin{equation}\label{02}
   \begin{array}{l}
     k_{1(2)}=\sqrt{2m(\varepsilon^f\pm h-\frac{\kappa^2}{2m} \pm i\omega \pm \frac{i}{2 \tau_{\uparrow(\downarrow)}})}, \\
      k_3=\sqrt{2m(\varepsilon^s-\frac{\kappa^2}{2m}+i\sqrt{\omega^2+\Delta^2})}, \\
   \end{array}
\end{equation}
where $k_{1(2)}$ corresponds to the majority (minority) spin
electron momentum in the left F layer for the antiparallel
configuration (at the same time in the right F layer with the
opposite magnetization direction majority and minority momenta we
label as $k_{4}$ and $k_{5}$), $k_{3}$ corresponds to the the S
layer electrons. For the parallel configuration $k_{1}=k_{4}$,
$k_{2}=k_{5}$. Here $\tau_{\uparrow(\downarrow)}$ is  the spin up
$(\uparrow)$ and spin down $(\downarrow)$ electron lifetime in the
ferromagnet, $1/\tau_{\uparrow(\downarrow)}=-2x_0 \gamma^2 Im
G^{dd}_{\uparrow\uparrow(\downarrow\downarrow)}$.

The procedure of DOS calculation reduces to the diagonalization of
the imaginary part of the normal Green function
\begin{equation}\label{DOS}
    N_{\uparrow(\downarrow)}(z)=\frac{1}{2 \pi}\int \kappa d\kappa
    G^{ss}_{\uparrow\uparrow(\downarrow\downarrow)}(z,z,i\omega \rightarrow \omega).
\end{equation}

Because of unhandiness of the expressions for the Green functions
here we cite as an example the normal Green function only of the
left F layer for spin up electrons:
\begin{equation}\label{G22}
    G^{ss}_{\uparrow\uparrow}(z,z',\omega)=f_1 e^{i k_1 z}
    +f_2 e^{-i k_1 z},
\end{equation}
where $f_1$ and $f_2$ are the functions of electron momenta,
$\omega$ and $\Delta$ which are given in the appendix.

\begin{figure}[h]
    \includegraphics[width=0.4\textwidth]{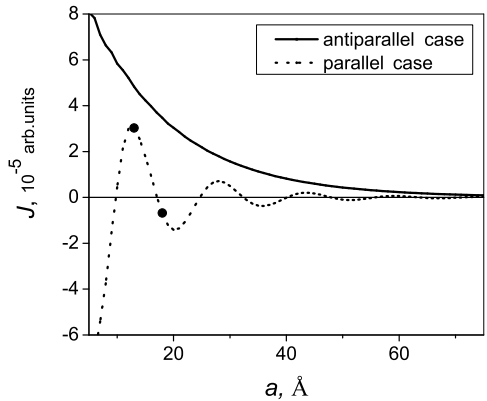}\\
 \caption{Josephson current $J$ versus F layers thickness $a$. Fermi momenta in F layer
  $k^{\uparrow}=1$ $\AA^{-1}$, $k^{\downarrow}=0.8$ $\AA^{-1}$
  ($k^{\uparrow(\downarrow)}=\sqrt{2m(\varepsilon^f\pm h}$);
 in S layer $k_{s}=1.2$ $\AA^{-1}$. Free path of electrons in F layer $l_{\uparrow}=100$
 $\AA$ and  $l_{\downarrow}=60$ $\AA$. The critical temperature of the superconducting
  metal is $T_c=4$ $K$.}\label{fig:Josephson}
\end{figure}

In order to discuss the connection between the behavior of DOS and
Josephson current which served as a motivation force of the
present study we rewrite here the expression for the current
derived earlier (see \cite{Pugach2004}) in the frame of Gorkov
equation for the antiparallel and parallel orientation of the
F-layer magnetic moments, correspondingly:
\begin{equation}\label{Jap}
    j_{AP}(0)=\frac{4e}{\pi}T\sum_\omega \int kdk \frac{(1-r^2)(1-R^2)sin\varphi}
    {(1-r^2)(1-R^2)cos\varphi+f_{AP}},
\end{equation}
\begin{equation}\label{Jp}
   j_P(0)=\frac{4e}{\pi}T\sum_\omega \int kdk Re\frac{(1-R^2)sin\varphi}
    {(1-R^2)cos\varphi+f_{P}},
\end{equation}
where $\varphi$ is the phase difference between right and left
superconducting electrodes order parameters, $R$ and $r$ are
reflection coefficients,
 $f_{AP}$ and $f_{P}$ are certain oscillatory
functions which depend on the ferromagnetic layers thickness $a$,
energy parameter $\omega$ and superconducting energy gap $\Delta$.
 There is no singularity in the Josephson current for the antiparallel
 case at low temperatures
 (see Fig. \ref{fig:Josephson}) predicted previously in
 \cite{BergeretLett2001,Golubov2002}.

\subsection{III. ANTIPARALLEL CASE}

\emph{Antiparallel case}. In Ref.\cite{Golubov2002} the authors
derive the expression for DOS which demonstrate the energy
renormalization due to exchange field in SFIFS (I - insulating
barier). For a certain exchange field value $h$ DOS expression
yields the singularity which is shifted to the Fermi level that
leads to the logarithmic divergency of the critical current at low
temperatures. However, the authors avoid the singularity problem
by taking the low barrier transparency which makes peaks broader
and in such a way suppresses the singularity.

In Ref.\cite{Koshina2000} energy variation of DOS was investigated
in the S/F structure. This case resembles more to the SFFS
structure with parallel magnetizations of F layers, but here we
can note that just as in Ref.\cite{Golubov2002}  DOS has a peak
and a region of zero value within the energy gap in much the same
way as in S/N structures ( N is the normal
metal)\cite{Golubov1995}.

\begin{figure}[h]
    \includegraphics[width=0.4\textwidth]{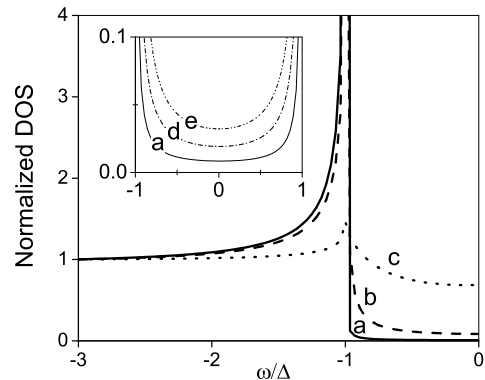}\\
    \caption{Energy variation of DOS at the left F/S border
 ($z=-a$) for the antiparallel case for the small value of
 exchange energy $h \sim \Delta$, $\Delta=1.4\, 10^{-3}eV$. a) $a=7$ $\AA$, $l^{\uparrow}=500$ $\AA$;
 b) $a=7$ $\AA$, $l_{\uparrow}=100$ $\AA$;
 c) $a=7$ $\AA$, $l_{\uparrow}=10$ $\AA$;
 d) $a=15$ $\AA$, $l_{\uparrow}=500$ $\AA$;
 e) $a=25$ $\AA$, $l_{\uparrow}=500$ $\AA$. }\label{fig:2}
\end{figure}

Both calculations were made for the small value of the exchange
field $h \sim \Delta$. Our computation for the same range of
exchange field is depicted in Fig. \ref{fig:2}. As one can see the
DOS has a BCS behavior outside the energy gap and usual divergence
at $\omega=\Delta$. However there in no any singularities or peaks
within the energy gap and DOS never equals to zero. Last
peculiarity may be explained by the destructive role of s-d
electron scattering in F layers which destroys the Cooper pairs.
The curves a,b,c were plotted for the different values of the free
path $l_{\uparrow(\downarrow)}$ of spin up(down) electrons in the
F layer, $l_{\uparrow(\downarrow)}=v_F
\tau_{\uparrow(\downarrow)}$, $v_F$ is the Fermi velocity in the F
layer. For the small values of free path (i.e. in the case of the
strong s-d scattering) the hollow becomes smaller and  DOS
approaches the constant as in the bulk ferromagnet. The inset of
Fig. \ref{fig:2} depicts energy DOS variation within the energy
gap for the different F layers thickness $a$. For the decreasing
values of $a$ DOS becomes more close to zero as in a bulk
superconductor.
\begin{figure}[h]   
  \includegraphics[width=0.4\textwidth]{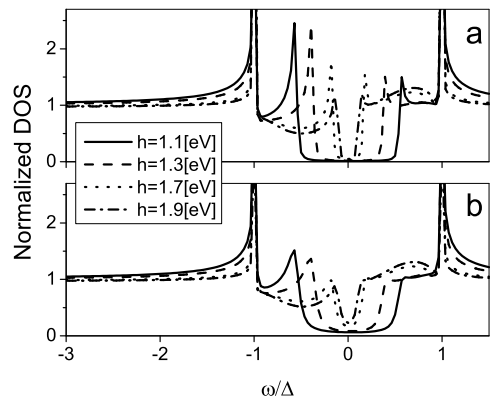}\\
  \caption{Energy variation of DOS at the F/S border ($z=a$)
  for the antiparallel alignment of
F-layers magnetization. Fermi momentum of electrons in S-layer:
$k_s=1$ $\AA^{-1}$. In the F-layer for the electrons with spin up:
$k^\uparrow=1$ $\AA^{-1}$,
$k^{\uparrow(\downarrow)}=\sqrt{2m(\varepsilon^f\pm h)}$. a)
$l_{\uparrow}=500$ $\AA$; b) $l_{\uparrow}=100$
$\AA$.}\label{fig:3}
\end{figure}
The peaks in DOS appear for the large values of the ferromagnetic
exchange field $h\sim 10^{3}\Delta$ (the case of a strong
ferromagnet). Outside the energy gap DOS stays of a BCS type and
within the gap DOS has two symmetrical peaks (Fig. \ref{fig:3}a).
As in Ref. \cite{Golubov2002} they can be shifted closer to the
Fermi energy ($\omega=0$) but at the same time their size
diminishes and there is no any singularity at the very zero. For
the smaller values of free path and large values of s-d scattering
parameter (Fig. \ref{fig:3}b) the peaks lose their pointed shape.

In both cases of small and large values of ferromagnetic exchange
field DOS does not demonstrate any divergence at the Fermi level
which is in a good  agreement with the calculated earlier
Josephson current dependencies for the antiparallel alignment of F
layers magnetic moments.

\subsection{IV. PARALLEL CASE}

\begin{figure}[h]   
  \includegraphics[width=0.5\textwidth]{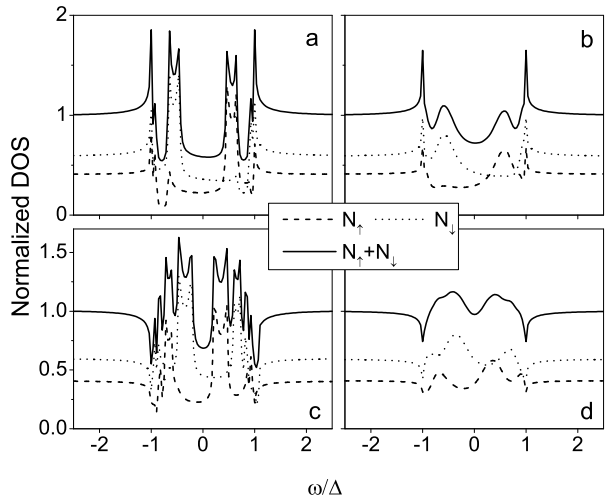}\\
  \caption{Energy variation of DOS at $z=0$
  for the parallel alignment of
F-layers magnetization. Fermi momentum of electrons in S-layer:
$k_s=1$ $\AA^{-1}$. In the F-layer for the electrons with spin up:
$k^\uparrow=1$ $\AA^{-1}$ and exchange field $h\sim 10^{3}
\Delta$. a) "0" state: $a=14$ $\AA$, $l_{\uparrow}=500$ $\AA$; b)
"0" state: $a=14$ $\AA$, $l_{\uparrow}=100$ $\AA$; c) "$\pi$"
state: $a=19$ $\AA$, $l_{\uparrow}=500$ $\AA$; d) "$\pi$" state:
$a=19$ $\AA$, $l^{\uparrow}=100$ $\AA$. Dashed line: DOS for spin
up $N_\uparrow$. Dotted line: DOS for spin down $N_\downarrow$.
Solid line: $N_\uparrow+N_\downarrow$.}\label{fig:4}
\end{figure}

 For the small values of the exchange field
$h\sim \Delta$ in the parallel case there are no any significant
changes in DOS in comparison with the antiparallel case presented
in Fig. \ref{fig:2}. But for the large values of $h$ in contrast
to the antiparallel case multiple electron reflection takes place
that graphically manifests itself in multiple resonance peaks in
DOS within the energy gap (see Fig. \ref{fig:4}).

This fact may be explained in the following way. The antiparallel
alignment is more favorable for the relatively free diffusion of
the Cooper pair electrons which have the opposite spin direction.
Possible destruction of the pair in the first F layer compensates
by the presence of the second F layer with the opposite magnetic
moment alignment and reversed majority and minority spins. While
in the parallel case for the large values of the exchange field
$h\gg \Delta$, when spin up and spin down electron energies differ
notably, only one electron in the pair whose Fermi energy is the
same in the F and S layers may travel relatively free. Hence, the
initial pair destroys rapidly and the multiple reflections becomes
possible as it happens in the potential well.

DOS dependencies in Fig. \ref{fig:4} are plotted for the different
values of free path. For increasing value of s-d scattering in the
F layers multiple irregularities are smoothed over and energy DOS
variations resemble to those calculated near the S/F border of
bilayer in \cite{Gusakova2004} in the frame of Gorkov equations.
So called "0" and "$\pi$" states are presented in Ref.
\cite{Kontos2001}. Due to the oscillations of the superconducting
order parameter induced in the F layers in the "$\pi$" state the
DOS shape is reversed with the respect to the normal "0" state.
Instead of usual BCS energy dependence with two peaks at $\omega
=\pm \Delta$ DOS has two dips at these energies values. As at was
discussed earlier the Josephson current varies its sign in the
parallel case \cite{Koshina2001,Golubov2002,Pugach2004}, ($+$)
corresponds to the "0" phase and ($-$) to the "$\pi$" phase. We
have compared DOS behavior with presented in Fig.
\ref{fig:Josephson} Josephson current oscillation from "0" to
"$\pi$" state in the parallel case. Thick points in Fig.
\ref{fig:Josephson} correspond to the F layers thicknesses $a$
which had been chosen for that comparison. Indeed, for these
values of $a$ DOS demonstrates the same transition between "0" and
"$\pi$" states (Fig. \ref{fig:4} a,b and Fig. \ref{fig:4} c,d,
correspondingly).

\subsection{V. SUMMARY}

 In this paper the energy DOS variation had been studied in the
SFFS structure for the antiparallel and parallel magnetic moments
alignment of two adjacent F layers. It was shown that there in no
any singularity in DOS in the antiparallel case that is in a good
agreement with calculated earlier Josephson current behavior in
the same model. In the parallel case the presence of "0" and
"$\pi$" states had been shown.

\subsection{ACKNOWLEDGMENTS}

We are grateful to N.Pugach, who had allowed us to cite here some
results (Eqs. (\ref{Jap}), (\ref{Jp}) and Fig.
\ref{fig:Josephson}) before their publication
(Ref.\cite{Pugach2004}). This work was supported by the Russian
Foundation for Basic Researches (grant N 04-02-16688a).

\subsection{APPENDIX: EXPRESSIONS FOR THE COEFFICIENTS IN EQ. (\ref{G22})}

\begin{widetext}
\begin{equation*}\label{app01}
  f_{1,2}=  \frac{e^{-i a k_4}}{4 k_1 k_4}(k_1\pm k_4) \left( A_1(k_3+k_4)+A_2(k_4-k_3^*) \right)+
\frac{e^{i a k_4}}{4 k_1 k_4}(k_1\mp k_4) \left(
A_1(k_4-k_3)+A_2(k_4+k_3^*) \right),
\end{equation*}
\begin{equation*}\label{app02}
\begin{array}{rcr}
 A_{1,2}& = & \frac{1}{den} \left[ \pm (B_2k_1-B_1k_3^*)(k_3t_{2,1}+k_2t_{4,3})
 \pm \Delta_2(B_1k_3+B_2k_1)(k_2t_{4,3}-k_3^*t_{2,1}) \pm \right.\\
& & \left. \pm
\Delta_1\Delta_2k_1(k_3+k_3^*)(B_2c_{2,1}-B_1c_{4,3})\right],\\
\end{array}
\end{equation*}
\begin{equation*}\label{app03}
\begin{array}{rcl}
den &= &
k_2(k_3+k_3^*)(t_2t_3-t_1t_4)+\Delta_1\Delta_2k_1(k_3+k_3^*)(c_1c_4+c_2c_3)+\\
& & +\Delta_1\left(
(k_1c_4-k_3^*c_2)(k_3t_1+k_2t_3)+(k_3^*c_1-k_1c_3)(k_3t_2+k_2t_4)\right)+\\
& & +\Delta_2\left(
(k_1c_4+k_3c_2)(k_2t_3-k_3^*t_1)+(k_3c_1-k_1c_3)(k_3^*t_2-k_2t_4)\right),\\
\end{array}
\end{equation*}
\begin{equation*}\label{app04}
\begin{array}{lr}
B_{1,2}=i(e^{-iz'k_1}e^{-iak_1}\mp e^{iz'k_1}e^{iak_1})/2k_1, &
\Delta_{1,2}=ie^{-i\varphi_1}\frac{\sqrt{\omega^2+\Delta^2}\pm i
\omega}{\Delta},
\end{array}
\end{equation*}
\begin{equation*}\label{app05}
\begin{array}{rcl}
  c_{1,3} & = & \frac{e^{-i a k_1}}{4k_1k_4}
  \left[ e^{-i a k_4}(k_3+k_4)(k_1+k_4)+ e^{i a k_4}(k_4-k_3)(k_1-k_4)\right] \pm\\
    &   & \pm \frac{e^{i a k_1}}{4k_1k_4}
  \left[ e^{-i a k_4}(k_3+k_4)(k_1-k_4)+ e^{i a k_4}(k_4-k_3)(k_1+k_4)\right], \\
  c_{2,4} & = & \frac{e^{-i a k_1}}{4k_1k_4}
  \left[ e^{-i a k_4}(k_4-k_3^*)(k_1+k_4)+ e^{i a k_4}(k_4+k_3^*)(k_1-k_4)\right] \pm\\
    &   & \pm \frac{e^{i a k_1}}{4k_1k_4}
  \left[ e^{-i a k_4}(k_4-k_3^*)(k_1-k_4)+ e^{i a k_4}(k_4+k_3^*)(k_1+k_4)\right], \\
\end{array}
\end{equation*}
\begin{equation*}\label{app06}
\begin{array}{rcl}
  t_{1,3} & = & \Delta_1\frac{e^{i a k_5}}{4k_2k_5}
  \left[ e^{i a k_2}(k_2-k_3)(k_2+k_5)+ e^{-i a k_2}(k_2+k_3)(k_5-k_2)\right] \pm\\
    &   & \pm \Delta_1\frac{e^{-i a k_5}}{4k_2k_5}
  \left[ e^{i a k_2}(k_2-k_3)(k_5-k_2)+ e^{-i a k_2}(k_2+k_3)(k_2+k_5)\right], \\
  t_{2,4} & = &-\Delta_2 \frac{e^{i a k_5}}{4k_2k_5}
  \left[ e^{i a k_2}(k_2+k_3^*)(k_2+k_5)+ e^{-i a k_2}(k_2-k_3^*)(k_5-k_2)\right] \mp\\
    &   & \mp \Delta_2 \frac{e^{-i a k_5}}{4k_2k_5}
  \left[ e^{i a k_2}(k_2+k_3^*)(k_5-k_2)+ e^{-i a k_4}(k_2-k_3^*)(k_2+k_5)\right]. \\
\end{array}
\end{equation*}
\end{widetext}


\begin{thebibliography}{99}

\bibitem{Buzdin1982}
A. I. Buzdin, L. N. Bulaevskii, S. V. Panjukov, Pis'ma Zh. Eksp.
Teor. Fiz. \textbf{35}, 147 (1982) [JETP Lett. \textbf{35}, 187
(1982)]; A. I. Buzdin, M. Yu. Kuprijanov, \emph{ibid.}
\textbf{53}, 308 (1991) [\emph{ibid.} \textbf{53}, 321 (1991)].

\bibitem{Kontos2001}
T. Kontos, M. Aprili, J. Lesueur, and X. Grison, Phys. Rev. Lett
\textbf{86}, 304 (2001).

\bibitem{Ryazanov2001}
V. V. Ryazanov \emph{et al.}, Phys. Rev. Lett \textbf{86}, 2427
(2001); A.V. Veretennikov \emph{et al.}, Physica B
\textbf{284-288}, 495 (2000).

\bibitem{BergeretLett2001}
F. S. Bergeret, A. F. Volkov, and K. B. Efetov, Phys. Rev.
Lett.\textbf{86}, 3140 (2001).

\bibitem{Koshina2001}
E. Koshina and V. Krivoruchko, Phys. Rev. B \textbf{63}, 224515
(2001), Phys. Rev. B \textbf{64}, 172511 (2001).

\bibitem{Golubov2002}
A. A. Golubov, M. Yu. Kupriyanov, Ya. V. Fominov, Sov. Phys.
Pis'ma v ZhETF \textbf{75}, iss. 3-4, 223 (2002).

\bibitem{Pugach2004}
N. Pugach, A. Vedyayev, J. Magn. Magn. Mater., to be published.

\bibitem{Koshina2000}
E. Koshina and V. Krivoruchko, Fiz.Nisk.Temp. \textbf{26}, 157
(2000).

\bibitem{Golubov1995}
A. A. Golubov \emph{et al.}, Phys.Rev. B \textbf{51}, 1073 (1995).

\bibitem{Gusakova2004}
D. Gusakova \emph{et al.}, J. Magn. Magn. Mater., to be published;
arXiv: cond-mat$\backslash 0401037$ (2004).


\end{thebibliography}
\end{document}